\documentclass[11pt,a4paper]{article}

\newif\ifanonymous
\anonymousfalse

\usepackage{amsmath}
\usepackage{amsfonts}
\usepackage{amssymb}
\usepackage{amsthm}
\usepackage[mathscr]{euscript}
\usepackage{bbm}
\usepackage{bm}
\usepackage{mathrsfs}
\usepackage{prodint}
\usepackage{graphicx}
\usepackage[left=0.7in,right=0.7in]{geometry}
\usepackage{setspace}
\usepackage{xcolor}
\usepackage{authblk}
\usepackage{hyperref}
\usepackage{paralist}

\graphicspath{{./figure/}}

\newcommand{\ourtitle}{{Non-Plug-In Estimators Could Outperform Plug-In Estimators:\\a Cautionary Note and a Diagnosis}}
\newcommand{\authoraffil}{{
\author{Hongxiang Qiu}
\affil{Department of Epidemiology and Biostatistics, Michigan State University}
}}
\newcommand{\acknowledge}{{
\section*{Acknowledgments}
I would like to express my gratitude to Dr. Zhehui Luo for her valuable comments.
This work was supported in part through computational resources and services provided by the Institute for Cyber-Enabled Research at Michigan State University.
}}

\newcommand{\blue}{\color{blue}}

\usepackage[round]{natbib}

\usepackage{algorithm}
\usepackage{algorithmicx}
\usepackage[noend]{algpseudocode}

\allowdisplaybreaks

\theoremstyle{definition}

\DeclareMathOperator{\logit}{logit}
\DeclareMathOperator{\expit}{expit}

\newcommand{\real}{{\mathbb{R}}}

\newcommand{\ind}{{\mathbbm{1}}}
\newcommand{\expect}{{\mathbb{E}}}

\newcommand{\naive}{{\mathrm{\text{na\"{i}ve}}}}
\newcommand{\TMLE}{\mathrm{TMLE}}
\newcommand{\DML}{{\mathrm{DML}}}

\title{\ourtitle}

\ifanonymous
\author{}
\else
\authoraffil
\fi

\date{}

\begin{document}

\maketitle

\begin{abstract}
    \textbf{Objectives:} Highly flexible nonparametric estimators have gained popularity in causal inference and epidemiology. Popular examples of such estimators include targeted maximum likelihood estimators (TMLE) and double machine learning (DML). TMLE is often argued or suggested to be better than DML estimators and several other estimators in small to moderate samples---even if they share the same large-sample properties---because TMLE is a plug-in estimator and respects the known bounds on the parameter, while other estimators might fall outside the known bounds and yield absurd estimates. However, this argument is not a rigorously proven result and may fail in certain cases.
    
    \textbf{Methods:} In a carefully chosen simulation setting, I compare the performance of several versions of TMLE and DML estimators of the average treatment effect among treated in small to moderate samples.
    
    \textbf{Results:} In this simulation setting, DML estimators outperforms some versions of TMLE in small samples. TMLE fluctuations are unstable, and hence empirically checking the magnitude of the TMLE fluctuation might alert cases where TMLE might perform poorly.
    
    \textbf{Conclusions:} As a plug-in estimator, TMLE is not guaranteed to outperform non-plug-in counterparts such as DML estimators in small samples. Checking the fluctuation magnitude might be a useful diagnosis for TMLE. More rigorous theoretical justification is needed to understand and compare the finite-sample performance of these highly flexible estimators in general.
\end{abstract}

\section{Introduction}

When estimating causal effects in epidemiology and causal inference, highly flexible nonparametric or semiparametric estimators have been increasingly popular.
Well-known examples of such estimators include augmented inverse probability weighting (AIPW) \citep{Robins1994,Robins1995}, one-step estimator \citep[e.g.,][]{Bickel1993,Pfanzagl1985}, targeted maximum likelihood (or minimum loss-based) estimators (TMLE) \citep{VanderLaan2006,VanderLaan2018}, and double/debiased machine learning (DML) \cite{Chernozhukov2017,Chernozhukov2018}.
\citet{Smith2023} found increasing popularity of TMLE, while the seminal DML paper \citet{Chernozhukov2018} has received more than 2600 citations as of this writing.
Compared to traditional parametric estimators, such highly flexible estimators minimize potential bias from making restrictive assumptions such as normality and linearity on the data-generating mechanism, coinciding with the culture of ``cautious causal inference'' \citep{Ogburn2021}.

One-step estimation, DML\footnote{AIPW is a special case of one-step estimation and DML for estimating the average treatment effect.} estimators and TMLE are both constructed based on semiparametric efficiency theory \citep[e.g.,][]{Pfanzagl1985,Pfanzagl1990,vandervaart1996} and share the same asymptotic (i.e., large-sample) behavior: They are all asymptotically normal with the same asymptotic variance.
For several estimands, including the average treatment effect and the average treatment effect among treated, these estimators require estimating two nuisance functions---the outcome regression and the propensity score---and are all asymptotically optimal with no further structural assumptions such as linearity, smoothness, or sparsity on these nuisance functions \citep{Balakrishnan2023,Jin2024}.
A unique property of TMLE compared to AIPW and DML estimators is that an appropriately constructed TMLE, as a plug-in estimator, respects known bounds on the parameter of interest, while AIPW and DML estimators, as non-plug-in estimators, might not. For example, if the parameter of interest is a probability such as a cumulative incidence, this parameter is known to lie between $0$ and $1$. An appropriately constructed TMLE takes the form of a sample analogue of a probability and thus must also lie between $0$ and $1$. In contrast, AIPW and DML estimators do not take the form of a probability, and so may be below $0$ or above $1$, violating the known bounds on a probability.

Because of the plug-in nature of TMLE, it has often been argued or suggested that TMLE is superior to non-plug-in estimators such as AIPW and DML estimators in small to moderate samples \citep[e.g.,][]{Gruber2010,Gruber2014,Levy2018CVTMLE,Tran2019,Rytgaard2021,Kennedy2022,Guo2023}.
This argument may appear intuitive, but it lacks theoretical justification.
Some of the aforementioned works have indeed found in numerical simulations that, when estimating the population average treatment effect or mean counterfactual outcomes, TMLE performs better than non-plug-in estimators near positivity violation, namely when the propensity score is close to zero or one.
However, these simulations do not deduce the superiority of TMLE in general.
The web post \citet{BlogTMLEvsOneStep} discussed general finite-sample behaviors of TMLE and other non-plug-in estimators with an indefinite conclusion.
The fact that an appropriately constructed TMLE respects known bounds does not logically preclude the possibility that it may perform worse than a non-plug-in estimator.

In this paper, I present a numerical counterexample of the above heuristic argument, showing that a TMLE respecting known bounds may perform worse than a non-plug-in estimator in small samples.
In this example, the estimand is the mean counterfactual outcome under control among treated, a key estimand for estimating the average treatment effect among treated (ATT), and the outcome is binary and rare.
Outcomes may be rare when studying rare diseases \citep[e.g.,][]{McCann2006,Barash2012} or constructing prediction intervals for individual treatment effects with high coverage \citep{Yang2022,Qiu2022}.
I also propose easy-to-implement methods to diagnose whether TMLE might yield poor finite-sample performance based on this example.

\section{Simulation setting}

\noindent\textbf{Causal estimand:} 
Let $X$ denote the baseline covariate, $A$ be the binary indicator of treatment ($A=1$) or control ($A=0$), and $Y$ be an observed outcome.
Let $Y(a)$ denote the potential outcome \citep{Neyman1923,Rubin1974} if, possibly counter to fact, the individual's treatment status should be set to $a \in \{0,1\}$.
The two potential outcomes $Y(1)$ and $Y(0)$ cannot both be observed.
Assume that $n$ independent and identically distributed copies of $(X,A,Y)$ are observed.
Consider the causal ATT estimand $\theta := \expect[Y(1) 
\mid A=1] - \expect[Y(0) \mid A=1]$. A closely related causal estimand is $\psi := \expect[Y(0) \mid A=1]$, the mean counterfactual outcome corresponding to control among treated.
Under standard causal assumptions---stable unit treatment value assumption (SUTVA), consistency, no unmeasured confounding, and positivity---$\psi$ can be identified as a summary $\expect[ \expect[Y \mid X,A=0] \mid A=1]$ of the observed data, and the ATT $\theta$ can be identified as $\expect[Y \mid A=1] - \psi$ \citep{Wang2017}.
I consider estimating $\theta$ by first constructing an efficient estimator of $\psi$, then estimating $\expect[Y \mid A=1]$ by the sample average of $Y$ among the treated, and then taking the difference between these two estimators. In the sequel, I will focus on describing methods to estimate $\psi$.

\noindent\textbf{Efficient influence function (EIF):} Define three nuisance parameters: the outcome regression function $Q(x):=\expect[Y \mid A=1,X=x]$, the propensity score function $g(x):=\Pr(A=1 \mid X=x)$ and the marginal probability of treatment $\pi:=\Pr(A=1)$.
Under a nonparametric model for the distribution of the observable data $(X,A,Y)$, the EIF for $\psi$ has been derived previously \citep{Hahn1998,VanderLaan2013}:
$$D(Q,g,\pi,\psi)(x,a,y) = \frac{\ind(a=0) g(x)}{\pi \{ 1-g(x) \}} \{y - Q(x)\} + \frac{\ind(a=1)}{\pi} \{Q(x) - \psi\},$$
that is, the asymptotic variance bound is $\expect[D(Q,g,\pi,\psi)(X,A,Y)^2]$.
The EIF takes a central role in constructing all aforementioned highly flexible efficient estimators because an asymptotically efficient estimator of $\psi$ must approximately equal $\psi + n^{-1} \sum_{i=1}^n D(Q,g,\pi,\psi)(X_i,A_i,Y_i)$, having an asymptotically normal distribution achieving the asymptotic variance bound.

\noindent\textbf{Data-generating process:} The data consists of $n$ independent and identically distributed observations consisting of covariate $X$, treatment indicator $A$, and outcome $Y$. I consider an ill scenario where the outcome $Y$ is binary and rare. I generate the covariate $X=(X^{(1)},X^{(2)},X^{(3)}) \in \real^3$ independently from $\mathrm{Unif}(-1,1)$, $A$ from $\mathrm{Bern}(\expit\{-1.4 + 0.1 X^{(1)} + 0.1 X^{(2)} - 0.1 X^{(3)}\})$, and $Y$ from $\mathrm{Bern}((1-A)\expit\{-4.64+(X^{(1)}+X^{(2)}+X^{(3)})/3\})$.
In other words, the treatment ($A=1$) eliminates the occurrence of the outcome.
The prevalence of treatment $\Pr(A=1)=19.8\%$.
The outcome is rare with prevalence in the control group being $\Pr(Y=1 \mid A=0)=1.01\%$.
I consider both small ($n=300$) and relatively large ($n=2000$) sample sizes.

For the small sample size, because of the low prevalence of the outcome, the number of cases in the data is often extremely small, a key feature of this simulation as a counterexample. Such a small number of cases may occur when studying rare diseases \citep[e.g.,][]{McCann2006,Barash2012}. A similar issue may also occur when estimating a low quantile of a continuous outcome, for example, when constructing prediction intervals of a counterfactual outcome or individual treatment effects with high coverage \citep{Yang2022,Qiu2022}, in which case the indicator of the outcome being below a threshold---a transformed binary outcome---needs to be considered.
Simulations with somewhat extreme sparsity are not uncommon in the literature \citep[e.g.,][]{Gruber2010}, and they are informative and useful.
Although extreme scenarios might be uncommon in practice, practitioners might not know \textit{a priori} whether the true data-generating mechanism is extreme and might hence wish to use methods that perform well even in extreme scenarios. Simulations with extreme setups shed light on methods' performance in such scenarios.

\noindent\textbf{Comparison between rare outcome and near positivity violation:} The ill scenario with rare outcome is in sharp contrast to another ill scenario with near positivity violation in previous simulation studies \citep[e.g.,][]{Gruber2010,Gruber2014,Tran2019}.
Positivity is nearly violated when the propensity score $\Pr(A=1 \mid X)$ is close to zero or one for some values of covariate $X$, yielding a lack of observations from either the treatment or the control group. 
This sparsity in treatment status further leads to a large asymptotic variance bound, reflecting the fundamentally large uncertainty due to the near violation of a crucial causal assumption.
On the contrary, when the outcome $Y$ is rare but the propensity score is bounded away from zero and one, no causal assumption is nearly violated. In fact, the asymptotic variance bound is small due to a lack of variability in the outcome. 
In both these ill scenarios, the asymptotic normal approximation to the estimator's distribution might be inaccurate in small to moderate samples because of the sparsity in one variable, whether it is the treatment status or the outcome.

\section{Methods compared in the simulation} \label{sec: methods}

I consider four versions of TMLE, all respecting the known bounds $[0,1]$ on $\psi$, and two versions of DML estimators. All these methods require estimating the nuisance functions $Q$ and $g$. I describe the main ideas of these methods in this section and present the algorithms in full detail in the supplemental material.

I use cross-fitting throughout to allow maximal flexibility in nuisance function estimation, which is made possible by the independence between the data training and evaluating nuisance function estimators because of sample-splitting \citep{Schick1986,Chernozhukov2018,Kennedy2022}.
Cross-fitting has also been empirically found to be superior to methods without sample-splitting \citep{Li2022}.
TMLE with cross-fitting is also termed cross-validated TMLE \citep{Zheng2011}.
The $n$ observations in the data are split into $V \geq 2$ disjoint folds of equal size. I choose $V=2$ for illustration. Let $I_v$ and $I_v^C$ denote the set of observation indices in and out of fold $v$, respectively.
For each fold $v$, the nuisance functions $Q$ and $g$ are estimated with data in $I_v^C$, namely out of fold $v$, as follows.
I use Super Learner \citep{VanderLaan2007}, a highly flexible ensemble learner, whose library consists of logistic regression, gradient boosting, random forest, and neural network with various tuning parameters to obtain an estimator $\hat{Q}_v$ of the outcome regression $Q$ \citep{Chen2016,Wright2017,Venables2002}.
The propensity score estimator $\hat{g}_v$ is estimated similarly except that, to alleviate overfitting, $\hat{g}_v$ is clipped to fall in the interval $[0.05,0.5]$ containing the true propensity score range $[0.15,0.25]$.
The marginal treatment probability estimator $\hat{\pi}_v$ of $\pi$ for fold $v$ is the sample proportion within fold $v$, namely $\hat{\pi}_v := \sum_{i \in I_v} \ind(A_i=1)/|I_v|$.
After constructing an estimator $\hat{\psi}_v$ of $\psi$ for fold $v$ using nuisance function estimators trained using data in $I_v^C$, the cross-fit estimator is the average of $\hat{\psi}_v$ over all folds.

I next briefly describe TMLE and DML estimators used to obtain an estimator for fold $v$.
Since the estimand is $\psi=\expect[Q(X) \mid A=1]$, it may seem natural to estimate $\psi$ by a sample analog $\hat{\psi}^{\naive}_v := \sum_{i \in I_v} \ind(A_i=1) \hat{Q}_v(X_i)/\sum_{i \in I_v} \ind(A_i=1)$, namely plugging in $\hat{Q}_v$ into the mean over observations in the treated group ($A=1$).
This na\"ive plug-in estimator is generally too biased and converges to the truth at suboptimal rates.
Its first-order bias is approximately $\sum_{i \in |I_v|} D(\hat{Q}_v,\hat{g}_v,\hat{\pi}_v,\hat{\psi}^{\naive}_v)(X_i,A_i,Y_i)/|I_v|$, a random quantity that can be computed from data.
Correcting this first-order bias is crucial to constructing an efficient estimator of $\psi$.

In TMLE, a fluctuation $\hat{Q}^*_v$ of $\hat{Q}_v$ is constructed carefully such that, with $\hat{\psi}^\TMLE_v := \sum_{i \in I_v} \ind(A_i=1) \hat{Q}^*_v(X_i)/\sum_{i \in I_v} \ind(A_i=1)$ being the plug-in estimator based on $\hat{Q}^*_v$, its first-order bias is zero, namely
\begin{equation}
    \frac{1}{|I_v|} \sum_{i \in |I_v|} D(\hat{Q}^*_v,\hat{g}_v,\hat{\pi}_v,\hat{\psi}^\TMLE_v)(X_i,A_i,Y_i) = \frac{1}{|I_v|} \sum_{i \in I_V} \frac{\ind(A_i=0) \hat{g}_v(X_i)}{ \hat{\pi}_v \{1-\hat{g}_v(X_i)\}} \left\{ Y_i - \hat{Q}^*_v(X_i) \right\}= 0. \label{eq: TMLE estimating equation}
\end{equation}
Construction of such a fluctuation is also called ``targeting''. As such, $\hat{\psi}^\TMLE_v$ has a small bias and is an efficient plug-in estimator. When $Y$ is known to be lie in $[0,1]$, as long as $\hat{Q}^*_v$ is ranged in $[0,1]$, the TMLE $\hat{\psi}^\TMLE_v$ also takes values in $[0,1]$, thus respecting known bounds on $\psi$.

In this simulation, I consider two types of fluctuations for TMLE.
\begin{enumerate}
    \item Clever covariate (\texttt{tmle\textunderscore c}): $\hat{Q}^*_v(x) = \expit \left\{ \logit \hat{Q}_v(x) + \epsilon^*_v \frac{\hat{g}_v(x)}{\hat{\pi}_v \{1-\hat{g}_v(x)\}} \right\}$, where $\epsilon^*_v$ is the fitted slope in the logistic regression with outcome $Y$, offset $\logit \hat{Q}_v(X)$, covariate $\frac{\hat{g}_v(x)}{\hat{\pi}_v \{1-\hat{g}_v(x)\}}$, and no intercept using observations with $A=0$ in fold $v$.

    \item Weighting (\texttt{tmle\textunderscore w}): $\hat{Q}^*_v(x) = \expit \left\{ \logit \hat{Q}_v(x) + \epsilon^*_v \right\}$, where $\epsilon^*_v$ is the fitted intercept in the intercept-only logistic regression with outcome $Y$, offset $\logit \hat{Q}_v(X)$, and weight $\frac{\ind(A=0) \hat{g}_v(X)}{\hat{\pi}_v \{1-\hat{g}_v(X)\}}$ using observations in fold $v$.
\end{enumerate}
In this simulation, I clip $\logit \hat{Q}_v(x)$ to fall in the interval $[10^{-4},10^4]$ to avoid numerical errors caused by $\hat{Q}_v(x)=0$. I do not clip the outcome regression $\hat{Q}_v(x)$ itself because, in practice when the outcome is rare, an investigator might not be able to preclude the possible existence of a covariate subgroup whose probability of $Y=1$ is exactly zero.
By construction, both fluctuated outcome regressions have a range in $[0,1]$, respecting known bounds on the outcome $Y$. Properties of logistic regression imply the desired small-bias property of $\hat{Q}^*_v$ in \eqref{eq: TMLE estimating equation} for both fluctuations \citep[see e.g., Section~6.5.1. in][]{Wakefield2013}.
\citet{Tran2023} claimed that, when estimating the average treatment effect under near positivity violation, a weighting approach is more robust than a clever covariate approach.
The generality of this phenomenon is still an open question.

\citet{Levy2018} proposed another implementation of cross-validated TMLE that runs only one regression over the entire data and thus only one fluctuation coefficient $\epsilon^*$. I also consider the two pooled regression variants, \texttt{tmle\textunderscore cp} and \texttt{tmle\textunderscore wp}, of \texttt{tmle\textunderscore c} and \texttt{tmle\textunderscore w}, respectively.

In contrast to TMLE, the DML estimator $\hat{\psi}^\DML_v$ directly solves the estimating equation for zero first-order bias:
$$\frac{1}{|I_v|} \sum_{i \in |I_v|} D(\hat{Q}_v,\hat{g}_v,\hat{\pi}_v,\hat{\psi}^{\DML}_v)(X_i,A_i,Y_i) = 0,$$
and is also efficient. This equation can be solved analytically to find that
$$\hat{\psi}^\DML_v = \hat{\psi}^{\naive}_v + \frac{1}{|I_v|} \sum_{i \in I_v} \frac{\ind(A_i=0) \hat{g}_v(X_i)}{\hat{\pi}_v \{ 1-\hat{g}_v(X_i) \}} \{ Y_i-\hat{Q}_v(X_i) \}.$$
The DML estimator $\hat{\psi}^\DML_v$ cannot be written as an average of an outcome regression model over the observations with $A_i=1$ and is thus a non-plug-in estimator.\footnote{The DML estimator can be viewed as an estimator that plugs in estimators of $Q$ and $g$ into the population EIF-based estimating equation in $\psi$, $\expect[D(Q,g,\pi,\psi)(X,A,Y)]=0$, which defines the estimand. However, in the TMLE literature, this plug-in view of the DML estimator is unconventional \citep[e.g.,][]{Gruber2010,Levy2018CVTMLE,Tran2019,Rytgaard2021,Kennedy2022,Guo2023}. Conventionally, a plug-in estimator considers (arguably) the most intuitive representation of the estimand in order to respect known bounds or restrictions on the estimand as well as possible. In contrast, the representation of $\psi$ based on EIF is arguably unintuitive. It might not respect known bounds on the estimand either.} This estimator may take values outside the interval $[0,1]$, even if the outcome $Y$ is known to fall in this interval and $\hat{Q}_v$ respects the known bounds.
The one-step estimator coincides with the DML estimator in this case.

In this simulation, besides the cross-fit DML estimator (\texttt{dml}), I also consider a post-processed version (\texttt{dml\textunderscore cl}) that clips the estimator to fall in the interval $[0,1]$, forcing the estimator to respect the known bounds on $\psi$.
For all methods, I also compute a 95\% Wald-confidence interval (CI) for $\psi$ based on a plug-in estimator of the EIF, namely the standard error takes the form of $\{ n^{-1} \sum_{i=1}^n D(\hat{Q},\hat{g},\hat{\pi},\hat{\psi})(X_i,A_i,Y_i)^2 \}^{-1/2}/\sqrt{n}$,
where $\hat{Q},\hat{g},\hat{\pi},\hat{\psi}$ are generic estimators of $Q,g,\pi,\psi$, respectively, depending on the estimator being used. Details can be found in the algorithms in the supplemental material.
CIs for $\theta$ can be constructed similarly. In this simulation, CIs for $\psi$ and $\theta$ differ by a sign because of no cases in the treated group ($A=1$).

I further consider a variant of TMLE developed for rare outcomes by \citet{Balzer2016}. This approach requires specifying relatively tight bounds on the outcome regression function $Q$, and uses a stabilized estimator of $Q$ by applying a linear transformation on the outcome to scale the range of $Q$ into the unit interval $[0,1]$ and using a logistic loss function to estimate $Q$. Specifying the bounds can be challenging, especially when strong effect modification is present, because the bounds are on a nuisance function $Q$ rather than an overall scalar summary of the population such as the ATT. Moreover, $Q$ is estimated by minimizing a logistic loss while the outcome may fall outside the unit interval, so fewer off-the-shelf machine learning methods are available for estimating $Q$ flexibly. In this simulation, when estimating and targeting $Q$, I consider two bounds on $Q$, $[0,0.05]$ and $[0,0.2]$, and use a scaled logistic model implemented via numerical optimization with manually coded objective functions and gradients as described in \citet{Balzer2016}. Because of these practical drawbacks compared to standard TMLE, the simulation will not prioritize this variant.
The code for the simulations is available at \url{https://github.com/QIU-Hongxiang-David/plug_in_sim}.

\section{Estimators' sampling distributions in the simulation} \label{sec: sim results}

\begin{figure}[bt]
    \centering
    \includegraphics[width=.9\linewidth]{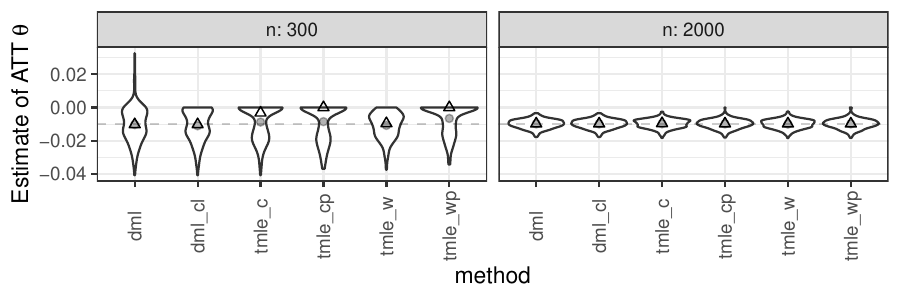}
    \caption{The six estimators' sampling distributions. The horizontal dashed line is the truth. The gray dots and black triangles are the means and medians, respectively.}
    \label{fig: est}
\end{figure}

I ran 200 simulation repetitions for each sample size.
The sampling distributions of the six estimators are presented in Fig.~\ref{fig: est} and Table~\ref{tab: bias MSE}.
In relatively large samples ($n=2000$), all estimators perform similarly: They are all asymptotically normal with the same efficient asymptotic variance, as indicated by asymptotic theory.
The Wald-CI coverage is somewhat low for all methods.
The relatively large sample size $n=2000$ appears not sufficiently large for the large-sample normal approximation to work well, but this sample size diminishes the difference across all the methods. Therefore, the low CI coverage is likely due to the binary nature of the outcome and the very low outcome prevalence, leading to challenges with normal approximation \citep[e.g.,][]{Brown2001}.

Before seeing the simulation results, since the truth $\psi=1.01\%$ is close to zero, one might conjecture that DML estimators perform worse than TMLE because of possibly violating known bounds in small samples ($n=300$).
Indeed, \texttt{dml} estimator is negative with a nontrivial probability (empirical proportion=10.5\%, 95\% CI: 7.0--15.5\%). 
However, as shown in Table~\ref{tab: bias MSE}, the bias of \texttt{dml} and its clipped version \texttt{dml\textunderscore cl} are both substantially smaller than all TMLEs except \texttt{tmle\textunderscore w}.
The mean squared error (MSE) of \texttt{dml} and \texttt{dml\textunderscore cl} does not appear superior due to larger variances.
The CIs based on \texttt{dml} and \texttt{dml\textunderscore cl} have substantially better coverage than all TMLEs except \texttt{tmle\textunderscore w}, although all CIs undercover.
As shown in Fig.~\ref{fig: est}, \texttt{tmle\textunderscore c}, \texttt{tmle\textunderscore cp}, and \texttt{tmle\textunderscore wp} all have a high probability of being close to zero, leading to an almost zero median. In other words, these three estimators overly underestimate too often.
Their sampling distributions also appear further from normal distribution than \texttt{dml}, \texttt{dml\textunderscore cl} and \texttt{tmle\textunderscore w}.
Therefore, in finite samples, TMLE---a plug-in estimator respecting known bounds on the parameter---may perform worse than DML---a non-plug-in estimator.

As a side note, the fold-wise weighted approach to TMLE, namely \texttt{tmle\textunderscore w}, appears to perform substantially better than the three other versions of TMLE in small samples, echoing \citet{Tran2023}. In fact, it performs the best among all methods regarding bias and MSE. 
The pooled TMLE approach proposed by \citet{Levy2018} appears suboptimal in small samples.

The sampling distributions of the TMLE variants proposed by \citet{Balzer2016} are presented in Fig.~\ref{fig: est small bound} and \ref{fig: est large bound} in the Supplemental Material.
All estimators perform similarly in relatively large samples ($n=2000$).
When the sample size is small ($n=300$) and the known bounds on $Q$ are relatively tight ($[0, 0.05]$), all versions of TMLE except \texttt{tmle\textunderscore w} have substantially improved performance and appear similar to the clipped DML estimator. The variant of \texttt{tmle\textunderscore w} performs worse as shown by a substantially increased bias and longer tail.
When the known bounds are much wider ($[0,0.2]$), these TMLE variants perform slightly worse as shown by longer tails in the sampling distributions, indicating some sensitivity of such TMLE variants to the specified bounds.
These results suggest that, when the outcome is rare and the sample size is relatively small, the variant developed by \citet{Balzer2016} might improve the performance of poorly-performing TMLE but harm well-performing TMLE.
It is still an open question whether a particular version of TMLE is superior to the other versions in general.

\begin{table}[bt]
    \centering
    \begin{tabular}{c|rrr|rrr}
    & \multicolumn{3}{c}{$n=300$} & \multicolumn{3}{|c}{$n=2000$} \\
    Method & Bias ($\times 10^{-4}$) & MSE ($\times 10^{-4}$) & CI coverage & Bias ($\times 10^{-4}$) & MSE ($\times 10^{-4}$) & CI coverage \\
    \hline
    \texttt{dml} & 9.78 & 1.18 & 83\% & -0.67 & 0.08 & 91\% \\
    \texttt{dml\textunderscore cl} & 2.80 & 0.96 & 83\% & -0.67 & 0.08 & 91\% \\
    \texttt{tmle\textunderscore c} & 22.78 & 0.91 & 67.5\% & 0.38 & 0.07 & 91.5\% \\
    \texttt{tmle\textunderscore cp} & 15.03 & 1.02 & 65.5\% & -0.93 & 0.09 & 91.5\% \\
    \texttt{tmle\textunderscore w} & -1.02 & 0.53 & 81.5\% & 0.04 & 0.07 & 91.5\% \\
    \texttt{tmle\textunderscore wp} & 29.9 & 0.86 & 65.5\% & -1.06 & 0.09 & 91.5\% \\
    \end{tabular}
    \caption{Monte Carlo estimates of the bias $\expect[\hat{\theta}-\theta]$, Mean Squared Error (MSE) $\expect[(\hat{\theta}-\theta)^2]$, and 95\% Wald-CI coverage $\Pr(\theta \in \mathrm{CI})$ of the six estimators.}
    \label{tab: bias MSE}
\end{table}

\section{Proposed diagnosis for TMLE} \label{sec: diagnosis}

\begin{figure}[bt]
    \centering
    \includegraphics[width=.9\linewidth]{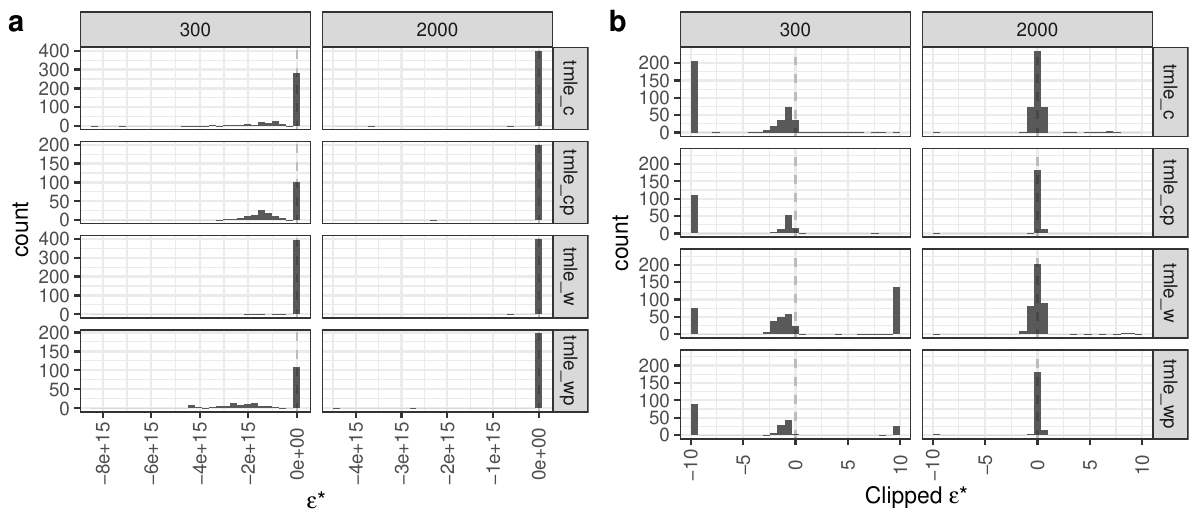}
    \caption{Histograms of the fluctuation coefficient $\epsilon^*_v$ or $\epsilon^*$ for the four versions of TMLE. Each column is a sample size and each row is a version of TMLE. The vertical dashed line represents no fluctuation, namely zero fluctuation coefficient. Subfigure~(b) presents $\epsilon^*_v$ or $\epsilon^*$ clipped to fall in the interval $[-10,10]$; the clipping affects only six values on samples with $n=2000$.}
    \label{fig: epsilon}
\end{figure}

To further investigate potential causes of the poorer small-sample performance of TMLE than DML estimators, I present the sampling distribution of the fluctuation coefficient $\epsilon^*_v$ of TMLE in Fig.~\ref{fig: epsilon}. 
Sample mean relative absolute differences (MRAD), $n^{-1} \sum_{v=1}^V \sum_{i \in I_v} \left| \hat{Q}_v^*(X_i)-\hat{Q}_v(X_i) /\hat{Q}_v^*(X_i) \right|$, between the fluctuation $\hat{Q}^*_v$ and the initial $\hat{Q}_v$ are presented in Fig.~\ref{fig: MRAD} in the supplemental material.
I choose the fluctuation $\hat{Q}_v^*$ rather than the initial $\hat{Q}_v$ as the denominator because $\hat{Q}_v(X_i)$ might equal zero exactly.
In Fig.~\ref{fig: Q} in the supplemental material, I also plot the predicted values of the fluctuated outcome regression $\hat{Q}^*_v$ in the simulated data set against those of the initial outcome regression $\hat{Q}_v$ (i) in a simulation run where the sample size is small ($n=300$), the poor-performing TMLEs are almost zero, and DML estimators is closer to the truth, and (ii) in a simulation run with a relatively large sample size ($n=2000$).

One sufficient condition to justify TMLE is that the initial outcome regression estimator $\hat{Q}_v$ is close to the truth $Q$ and the magnitude of fluctuation is small.
It is plausible that the fluctuation has a small magnitude in large samples because the targeting step refits the outcome regression based on a well-trained initial fit.
Indeed, this appears to hold approximately in relatively large samples ($n=2000$), with $\epsilon^*_v$ and sample mean MRAD concentrating around zero (Fig.~\ref{fig: epsilon}b, Fig.~\ref{fig: MRAD}b), and $\hat{Q}^*_v(X_i)$ similar to $\hat{Q}_v(X_i)$ (Fig.~\ref{fig: Q} Row~2).
However, in small samples ($n=300$), there is a high probability that the fluctuation has a large magnitude, namely $\epsilon^*_v$ and sample MRAD has huge magnitudes (Fig.~\ref{fig: epsilon}a, Fig.~\ref{fig: MRAD}a) and $\hat{Q}^*_v(X_i)$ is far from $\hat{Q}_v(X_i)$ (Fig.~\ref{fig: Q} Row~1).
In this simulation, because of clipping $\logit \hat{Q}_v(x)$ into a large finite interval to account for $\hat{Q}_v(x)=0$, one scenario of a significant fluctuation is no cases in the data, although other scenarios also exist.
In the simulation run shown in Fig.~\ref{fig: Q} Row~1, at least one fluctuation has a large magnitude (of order $10^3$ or even $10^{15}$) for all four versions of TMLEs.

Based on these simulation results, I conjecture that TMLE might perform unsatisfactorily in a given data if the magnitude $|\epsilon^*_v|$ of the fluctuation is large and if the predicted values of the fluctuated and initial outcome regressions are qualitatively different.

Therefore, it might be beneficial to check at least one of the following:
\begin{itemize}
    \item For each fold $v$, after computing the fluctuated outcome regression estimator $\hat{Q}^*_v$, typically by fitting a generalized linear model with fitted coefficient $\epsilon^*_v$, if $|\epsilon^*_v|$ is extremely large (e.g., greater than 10), the TMLE might perform unsatisfactorily. When a pooled logistic regression is used for targeting, if the one fluctuation coefficient $\epsilon^*$ has an extremely large absolute value, the TMLE might perform unsatisfactorily.
    \item For every observation index $i$, compute the predicted value $\hat{Q}_i^* := \hat{Q}_v^*(X_i)$ and $\hat{Q}_i := \hat{Q}_v(X_i)$ for the fluctuated and initial outcome regression estimators, where $v$ is the fold containing observation $i$ (i.e., $i \in I_v$). Compute the sample MRAD, $n^{-1} \sum_{i=1}^n \left| \hat{Q}_i^*-\hat{Q}_i /\hat{Q}_i^* \right|$. If MRAD is extremely large (e.g., greater than 10), the TMLE might perform unsatisfactorily.
    \item Plot $\hat{Q}_i^*$ against $\hat{Q}_i$ for all observations in the data, for example, as Fig.~\ref{fig: Q} in the supplemental material. If many points appear far from the diagonal line, this indicates discordance between $\hat{Q}^*_v$ and $\hat{Q}_v$ and the TMLE might perform unsatisfactorily. Compared to $|\epsilon^*_v|$ and MRAD, this plot provides a more intuitive visualization of the concordance between $\hat{Q}^*_v$ and $\hat{Q}_v$.
\end{itemize}
The above thresholds for large $|\epsilon^*_v|$ and MRAD are chosen based on this simulation and are somewhat arbitrary.

\begin{figure}[bt]
    \centering
    \includegraphics[width=\linewidth]{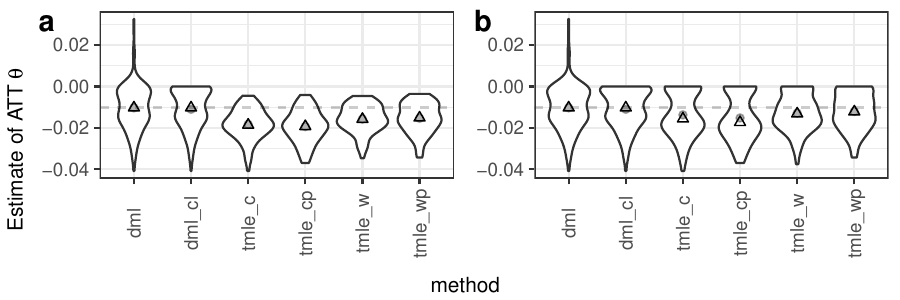}
    \caption{Plots similar to Fig.~\ref{fig: est} after removing potentially problematic TMLE in small samples ($n=300$). In Subfigures~(a) and (b), TMLE estimates with $|\epsilon^*_v|>10$ and sample MRAD greater than 10 are removed, respectively.}
    \label{fig: check diagnosis}
\end{figure}

All these diagnoses can be easily implemented because they use intermediate computation results from TMLE.
It remains challenging to identify a clear theoretically justified threshold to determine whether the fluctuation is too large or acceptable.
After removing TMLE estimates satisfying one of these conditions, the TMLE sampling distributions appear better (see Fig.~\ref{fig: check diagnosis}).
However, since excluding estimators based on diagnosis tools is conditioning on a subset of all realizations of the data, statistical properties of the procedure such as bias and CI coverage are not guaranteed by theory. As shown in Table~\ref{tab: check diagnosis}, after excluding simulation runs based on the above diagnosis, for all estimators, the bias has become larger. The Wald-CI overcovers or undercover, but the coverage has improved when it undercovers. The MSE appears to remain similar. How to incorporate diagnosis into statistical procedures with theoretical guarantees is an interesting open question.

I conclude by remarking that TMLE might still perform well even if the fluctuation is large, for example, \texttt{tmle\textunderscore w} in this simulation.
Thus, a large fluctuation should be interpreted as a warning rather than a falsification of TMLE.

\begin{figure}[bt]
    \centering
    \includegraphics[width=\linewidth]{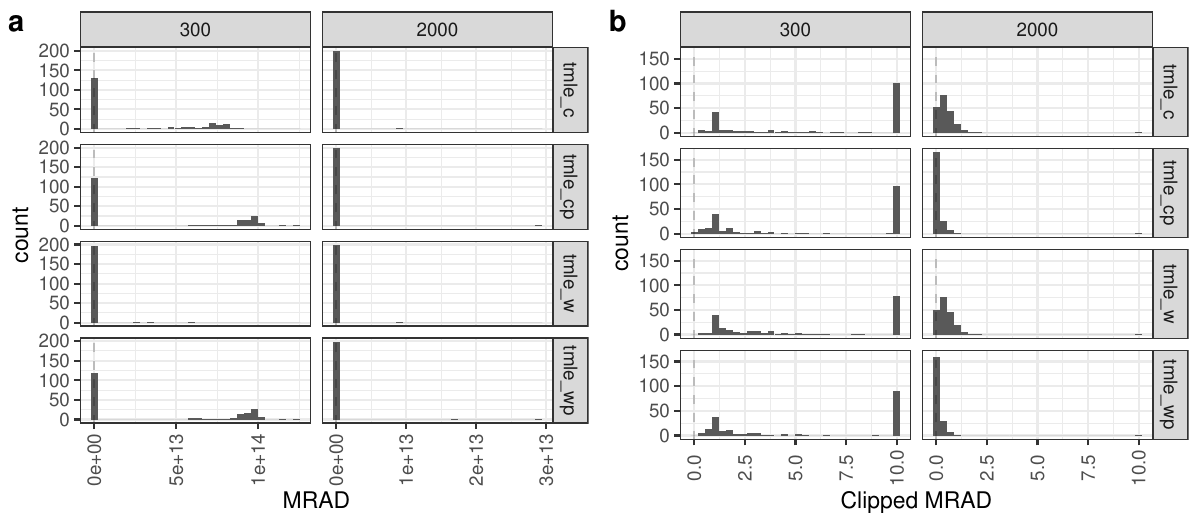}
    \caption{Sampling distributions of sample MRAD. Subfigure~(b) presents MRAD clipped to fall in the interval $[0,10]$; the clipping affects only six values in all simulation runs with $n=2000$.}
    \label{fig: MRAD}
\end{figure}

\begin{table}[bt]
    \centering
    \begin{tabular}{c|rrr|rrr}
    & \multicolumn{3}{c}{Remove $|\epsilon^*_v|>10$} & \multicolumn{3}{|c}{Remove $|\text{MRAD}|>10$} \\
    Method & Bias ($\times 10^{-4}$) & MSE ($\times 10^{-4}$) & CI coverage & Bias ($\times 10^{-4}$) & MSE ($\times 10^{-4}$) & CI coverage \\
    \hline
    \texttt{dml} & -46.22 & 1.05 & 100\% & -17.28 & 0.98 & 81.1\% \\
    \texttt{dml\textunderscore cl} & -48.64 & 0.97 & 100\% & -18.81 & 0.93 & 81.1\% \\
    \texttt{tmle\textunderscore c} & -78.79 & 1.19 & 98.9\% & -37.58 & 1.14 & 79.5\% \\
    \texttt{tmle\textunderscore cp} & -81.81 & 1.45 & 96.8\% & -37.60 & 1.35 & 76.2\% \\
    \texttt{tmle\textunderscore w} & -59.23 & 0.85 & 100\% & -26.53 & 0.85 & 81.1\% \\
    \texttt{tmle\textunderscore wp} & -37.30 & 0.84 & 96.8\% & -6.45 & 0.86 & 76.2\% \\
    \end{tabular}
    \caption{Table similar to Table.~\ref{tab: bias MSE} after removing potentially problematic TMLE in small samples ($n=300$).}
    \label{tab: check diagnosis}
\end{table}

\section{Discussion}

I show in a simulation that in finite samples, a plug-in estimator might perform worse than a non-plug-in estimator sharing the same large-sample properties, even though the plug-in estimator always respects known bounds on the parameter while the non-plug-in estimator might not.

This example is cherry-picked and has several limitations.
In practice, when the outcome is rare, other sampling schemes such as case-control are often adopted to overcome the challenge of too few cases. However, a preliminary analysis with data from a small cohort or cross-sectional study might fit this example. 
If the propensity score is estimated with a flexible machine learning algorithm that might overfit in small samples (e.g., without clipping), the propensity score estimator is more vulnerable to overfitting and near positivity violation, and thus DML estimators may perform much worse than TMLE in small samples.
If the estimand is the ATT instead of the mean counterfactual outcome among treated, TMLE might also perform well due to the small magnitude of its bias.
Nevertheless, the mean counterfactual outcome may still be of scientific interest because it informs the counterfactual result of one intervention. It is also closely related to covariate shift problems in machine learning \citep{Lei2021,Qiu2022}. 

I use TMLE and DML estimators as examples from the two categories of estimators---plug-in and non-plug-in---but counterexamples might exist for other pairs of estimators with the same large-sample behaviors.
I further propose methods to diagnose whether TMLE might perform worse than DML estimators due to a large fluctuation based on this simulation.
Because intuitive arguments such as the plug-in principle may be flawed, more rigorous theoretical justification is needed to understand and compare the finite-sample performance of highly flexible estimators in general.

\ifanonymous
\else
\acknowledge
\fi

\bibliographystyle{abbrvnat}
\bibliography{ref}

\clearpage

\setcounter{page}{1}
\setcounter{section}{0}
\renewcommand{\thesection}{S\arabic{section}}%
\setcounter{table}{0}
\renewcommand{\thetable}{S\arabic{table}}%
\setcounter{figure}{0}
\renewcommand{\thefigure}{S\arabic{figure}}%
\setcounter{equation}{0}
\renewcommand{\theequation}{S\arabic{equation}}%
\setcounter{lemma}{0}
\renewcommand{\thelemma}{S\arabic{lemma}}%
\setcounter{theorem}{0}
\renewcommand{\thetheorem}{S\arabic{theorem}}%
\setcounter{corollary}{0}
\renewcommand{\thecorollary}{S\arabic{corollary}}%
\setcounter{algorithm}{0}
\renewcommand{\thealgorithm}{S\arabic{algorithm}}%

\begin{center}
    \LARGE Supplement to ``\ourtitle''
\end{center}

I present the six estimators of the mean counterfactual outcome corresponding to control among the treated, $\expect[Y(0) \mid A=1]$, described in Section~\ref{sec: methods}. The major differences (i) among the four TMLEs, and (ii) between the two DML estimators, are highlighted in blue. Methods to estimate the nuisance functions are described in Section~\ref{sec: methods}. The method to compute standard errors and construct Wald-CI is non-unique; only one simple working method is described. I use $z_\alpha$ to denote the $(1-\alpha/2)$-quantile of the standard normal distribution. After obtaining an estimator $\hat{\psi}$ of $\psi$, the ATT $\theta$ is estimated by
$$\hat{\theta} := \frac{\sum_{i=1}^n \ind(A_i=1) Y_i}{\sum_{i=1}^n \ind(A_i=1)} - \hat{\psi}.$$

\begin{algorithm}
    \caption{Estimator \texttt{tmle\textunderscore c} of $\psi=\expect[Y(0) \mid A=1]$.}
    \begin{algorithmic}[1]
        \State Randomly split data into $V$ folds of approximately equal sizes. Denote the index sets of observations in and out of fold $v$ by $I_v$ and $I_v^C$, respectively.
        \For{$v = 1, \ldots, V$}
            \State Estimate the outcome regression $Q$ by $\hat{Q}_v$ trained using data in the control group ($A_i=0$) in $I_v^C$ (namely outside fold $v$).
            \State Estimate the propensity score $g$ by $\hat{g}_v$ trained using data in $I_v^C$ (namely outside fold $v$).
            \State Estimate the marginal probability of treatment $\pi$ by $\hat{\pi}_v := \sum_{i \in I_v} \ind(A_i=1)/|I_v|$.
            {\blue
            \State Define $H_v(x):=\frac{\hat{g}_v(X)}{\hat{\pi}_v \{1-\hat{g}_v(X)\}}$. \Comment{(Define clever covariate)}
            \State Using observations in $I_v$ in the control group ($A_i=0$), run a logistic regression with outcome $Y_i$, offset $\logit \hat{Q}_v(X_i)$, covariate $H_v(X_i)$, and no intercept. Denote the fitted slope by $\epsilon^*_v$. \Comment{(Targeting within each fold)}
            \State Define the targeted outcome regression $\hat{Q}^*_v(x) := \expit \left\{ \logit \hat{Q}_v(x) + \epsilon^*_v H_v(x) \right\}$.
            \State Compute the plug-in estimator based on the targeted outcome regression, namely
            $$\hat{\psi}_v := \frac{\sum_{i \in I_v} \ind(A_i=1) \hat{Q}^*_v(X_i)}{\sum_{i \in I_v} \ind(A_i=1)}.$$
            }
        \EndFor
        \blue
        \State Output the cross-fit estimator $\hat{\psi} = n^{-1} \sum_{v=1}^V |I_v| \hat{\psi}_v$.
        \State Wald-CI: Compute the standard error $\mathrm{SE} := \{ n^{-1} \sum_{v=1}^V \sum_{i \in I_v} D(\hat{Q}_v^*,\hat{g}_v,\hat{\pi}_v,\hat{\psi})(X_i,A_i,Y_i)^2 \}^{-1/2}/\sqrt{n}$ and construct two-sided $(1-\alpha)$-level Wald-CI $\hat{\psi} \pm z_\alpha \times \mathrm{SE}$.
    \end{algorithmic}
\end{algorithm}

\begin{algorithm}
    \caption{Estimator \texttt{tmle\textunderscore w} of $\psi=\expect[Y(0) \mid A=1]$.}
    \begin{algorithmic}[1]
        \State Randomly split data into $V$ folds of approximately equal sizes. Denote the index sets of observations in and out of fold $v$ by $I_v$ and $I_v^C$, respectively.
        \For{$v = 1, \ldots, V$}
            \State Estimate the outcome regression $Q$ by $\hat{Q}_v$ trained using data in the control group ($A_i=0$) in $I_v^C$ (namely outside fold $v$).
            \State Estimate the propensity score $g$ by $\hat{g}_v$ trained using data in $I_v^C$ (namely outside fold $v$).
            \State Estimate the marginal probability of treatment $\pi$ by $\hat{\pi}_v := \sum_{i \in I_v} \ind(A_i=1)/|I_v|$.
            {\blue
            \State Define $W_v(a,x):=\frac{\ind(a=0) \hat{g}_v(x)}{\hat{\pi}_v \{1-\hat{g}_v(x)\}}$. \Comment{(Define weight)}
            \State Using data in $I_v$, run an intercept-only logistic regression with outcome $Y_i$, offset $\logit \hat{Q}_v(X_i)$, and weight $W_v(A_i,X_i)$. Denote the fitted intercept by $\epsilon^*_v$. \Comment{(Targeting within each fold)}
            \State Define the targeted outcome regression $\hat{Q}^*_v(x) := \expit \left\{ \logit \hat{Q}_v(x) + \epsilon^*_v \right\}$.
            \State Compute the plug-in estimator based on the targeted outcome regression, namely
            $$\hat{\psi}_v := \frac{\sum_{i \in I_v} \ind(A_i=1) \hat{Q}^*_v(X_i)}{\sum_{i \in I_v} \ind(A_i=1)}.$$
            }
        \EndFor
        \blue
        \State Output the cross-fit estimator $\hat{\psi} = n^{-1} \sum_{v=1}^V |I_v| \hat{\psi}_v$.
        \State Wald-CI: Compute the standard error $\mathrm{SE} := \{ n^{-1} \sum_{v=1}^V \sum_{i \in I_v} D(\hat{Q}_v^*,\hat{g}_v,\hat{\pi}_v,\hat{\psi})(X_i,A_i,Y_i)^2 \}^{-1/2}/\sqrt{n}$ and construct two-sided $(1-\alpha)$-level Wald-CI $\hat{\psi} \pm z_\alpha \times \mathrm{SE}$.
    \end{algorithmic}
\end{algorithm}

\begin{algorithm}
    \caption{Estimator \texttt{tmle\textunderscore cp} of $\psi=\expect[Y(0) \mid A=1]$.}
    \begin{algorithmic}[1]
        \State Randomly split data into $V$ folds of approximately equal sizes. Denote the index sets of observations in and out of fold $v$ by $I_v$ and $I_v^C$, respectively.
        \For{$v = 1, \ldots, V$}
            \State Estimate the outcome regression $Q$ by $\hat{Q}_v$ trained using data in the control group ($A_i=0$) in $I_v^C$ (namely outside fold $v$).
            \State Estimate the propensity score $g$ by $\hat{g}_v$ trained using data in $I_v^C$ (namely outside fold $v$).
            \State Estimate the marginal probability of treatment $\pi$ by $\hat{\pi}_v := \sum_{i \in I_v} \ind(A_i=1)/|I_v|$.
            {\blue
            \State For each $i \in I_v$, define $\hat{Q}_i := \hat{Q}_v(X_i)$ and $H_i := \frac{\hat{g}_v(X_i)}{\hat{\pi}_v \{1-\hat{g}_v(X_i)\}}$. \Comment{(Define clever covariate)}
            }
        \EndFor
        \blue
        \State Using observations in the control group ($A=0$) in the entire dataset, run a logistic regression with outcome $Y_i$, offset $\logit \hat{Q}_v(X)$, covariate $H_v(X)$, and no intercept. Denote the fitted slope by $\epsilon^*$. \Comment{(Targeting over the entire, namely pooled, data)}
        \State Compute the targeted outcome regression evaluated at each observation $\hat{Q}^*_i := \expit \left\{ \logit \hat{Q}_i + \epsilon^* H_i \right\}$.
        \State Output the plug-in estimator based on the targeted outcome regression
        $$\hat{\psi} := \frac{\sum_{i=1}^n \ind(A_i=1) \hat{Q}^*_i}{\sum_{i=1}^n \ind(A_i=1)}.$$
        \State Wald-CI: With $\hat{Q}_v^* := \expit \left\{ \logit \hat{Q} + \epsilon^* \frac{\hat{g}_v}{\hat{\pi}_v \{1-\hat{g}_v\}} \right\}$, compute the standard error $\mathrm{SE} := \{ n^{-1} \sum_{v=1}^V \sum_{i \in I_v} D(\hat{Q}_v^*,\hat{g}_v,\hat{\pi}_v,\hat{\psi})(X_i,A_i,Y_i)^2 \}^{-1/2}/\sqrt{n}$ and construct two-sided $(1-\alpha)$-level Wald-CI $\hat{\psi} \pm z_\alpha \times \mathrm{SE}$.
    \end{algorithmic}
\end{algorithm}

\begin{algorithm}
    \caption{Estimator \texttt{tmle\textunderscore wp} of $\psi=\expect[Y(0) \mid A=1]$.}
    \begin{algorithmic}[1]
        \State Randomly split data into $V$ folds of approximately equal sizes. Denote the index sets of observations in and out of fold $v$ by $I_v$ and $I_v^C$, respectively.
        \For{$v = 1, \ldots, V$}
            \State Estimate the outcome regression $Q$ by $\hat{Q}_v$ trained using data in the control group ($A_i=0$) in $I_v^C$ (namely outside fold $v$).
            \State Estimate the propensity score $g$ by $\hat{g}_v$ trained using data in $I_v^C$ (namely outside fold $v$).
            \State Estimate the marginal probability of treatment $\pi$ by $\hat{\pi}_v := \sum_{i \in I_v} \ind(A_i=1)/|I_v|$.
            {\blue
            \State For each $i \in I_v$, define $\hat{Q}_i := \hat{Q}_v(X_i)$ and $W_i := \frac{\ind(A_i=0) \hat{g}_v(X_i)}{\hat{\pi}_v \{1-\hat{g}_v(X_i)\}}$. \Comment{(Define weight)}
            }
        \EndFor
        \blue
        \State Using observations in the entire dataset, run an intercept-only logistic regression with outcome $Y_i$, offset $\logit \hat{Q}_i$, and weight $W_i$. Denote the fitted intercept by $\epsilon^*$. \Comment{(Targeting over the entire, namely pooled, data)}
        \State Compute the targeted outcome regression evaluated at each observation $\hat{Q}^*_i := \expit \left\{ \logit \hat{Q}_i + \epsilon^* \right\}$.
        \State Compute the plug-in estimator based on the targeted outcome regression
        $$\hat{\psi} := \frac{\sum_{i=1}^n \ind(A_i=1) \hat{Q}^*_i}{\sum_{i=1}^n \ind(A_i=1)}.$$
        \State Wald-CI: With $\hat{Q}_v^* := \expit \left\{ \logit \hat{Q} + \epsilon^* \right\}$, compute the standard error $\mathrm{SE} := \{ n^{-1} \sum_{v=1}^V \sum_{i \in I_v} D(\hat{Q}_v^*,\hat{g}_v,\hat{\pi}_v,\hat{\psi})(X_i,A_i,Y_i)^2 \}^{-1/2}/\sqrt{n}$ and construct two-sided $(1-\alpha)$-level Wald-CI $\hat{\psi} \pm z_\alpha \times \mathrm{SE}$.
    \end{algorithmic}
\end{algorithm}

\begin{algorithm}
    \caption{Estimator \texttt{dml} of $\psi=\expect[Y(0) \mid A=1]$.}
    \begin{algorithmic}[1]
        \State Randomly split data into $V$ folds of approximately equal sizes. Denote the index sets of observations in and out of fold $v$ by $I_v$ and $I_v^C$, respectively.
        \For{$v = 1, \ldots, V$}
            \State Estimate the outcome regression $Q$ by $\hat{Q}_v$ trained using data in the control group ($A_i=0$) in $I_v^C$ (namely outside fold $v$).
            \State Estimate the propensity score $g$ by $\hat{g}_v$ trained using data in $I_v^C$ (namely outside fold $v$).
            \State Estimate the marginal probability of treatment $\pi$ by $\hat{\pi}_v := \sum_{i \in I_v} \ind(A_i=1)/|I_v|$.
            \State Compute the DML estimator
            $$\hat{\psi}_v := \frac{1}{|I_v|} \sum_{i \in I_v} \left\{ \frac{\ind(A_i=1) \hat{Q}_v(X_i)}{\ind(A_i=1)} + \frac{\ind(A_i=0) \hat{g}_v(X_i)}{\hat{\pi}_v \{1-\hat{g}_v(X_i)\}} \{ Y_i-\hat{Q}_v(X_i) \} \right\}.$$
        \EndFor
        \State Output the cross-fit DML estimator $\hat{\psi} = n^{-1} \sum_{v=1}^V |I_v| \hat{\psi}_v$.
        \State Wald-CI: Compute the standard error $\mathrm{SE} := \{ n^{-1} \sum_{v=1}^V \sum_{i \in I_v} D(\hat{Q}_v,\hat{g}_v,\hat{\pi}_v,\hat{\psi})(X_i,A_i,Y_i)^2 \}^{-1/2}/\sqrt{n}$ and construct two-sided $(1-\alpha)$-level Wald-CI $\hat{\psi} \pm z_\alpha \times \mathrm{SE}$.
    \end{algorithmic}
\end{algorithm}

\begin{algorithm}
    \caption{Estimator \texttt{dml\textunderscore cl} of $\psi=\expect[Y(0) \mid A=1]$.}
    \begin{algorithmic}[1]
        \State Randomly split data into $V$ folds of approximately equal sizes. Denote the index sets of observations in and out of fold $v$ by $I_v$ and $I_v^C$, respectively.
        \For{$v = 1, \ldots, V$}
            \State Estimate the outcome regression $Q$ by $\hat{Q}_v$ trained using data in the control group ($A_i=0$) in $I_v^C$ (namely outside fold $v$).
            \State Estimate the propensity score $g$ by $\hat{g}_v$ trained using data in $I_v^C$ (namely outside fold $v$).
            \State Estimate the marginal probability of treatment $\pi$ by $\hat{\pi}_v := \sum_{i \in I_v} \ind(A_i=1)/|I_v|$.
            \State Compute the DML estimator
            $$\hat{\psi}_v := \frac{1}{|I_v|} \sum_{i \in I_v} \left\{ \frac{\ind(A_i=1) \hat{Q}_v(X_i)}{\ind(A_i=1)} + \frac{\ind(A_i=0) \hat{g}_v(X_i)}{\hat{\pi}_v \{1-\hat{g}_v(X_i)\}} \{ Y_i-\hat{Q}_v(X_i) \} \right\}.$$
        \EndFor
        \State Compute the cross-fit DML estimator $\hat{\psi} = n^{-1} \sum_{v=1}^V |I_v| \hat{\psi}_v$.
        \blue
        \State Output $\hat{\psi}^\dagger := \min \{ \max\{\hat{\psi},0\}, 1\}$. \Comment{(Clipping to $[0,1]$)}
        \State Wald-CI: Compute the standard error $\mathrm{SE} := \{ n^{-1} \sum_{v=1}^V \sum_{i \in I_v} D(\hat{Q}_v,\hat{g}_v,\hat{\pi}_v,\hat{\psi}^\dagger)(X_i,A_i,Y_i)^2 \}^{-1/2}/\sqrt{n}$ and construct two-sided $(1-\alpha)$-level Wald-CI $\hat{\psi}^\dagger \pm z_\alpha \times \mathrm{SE}$.
    \end{algorithmic}
\end{algorithm}

\begin{figure}[bt]
    \centering
    \includegraphics[width=.9\linewidth]{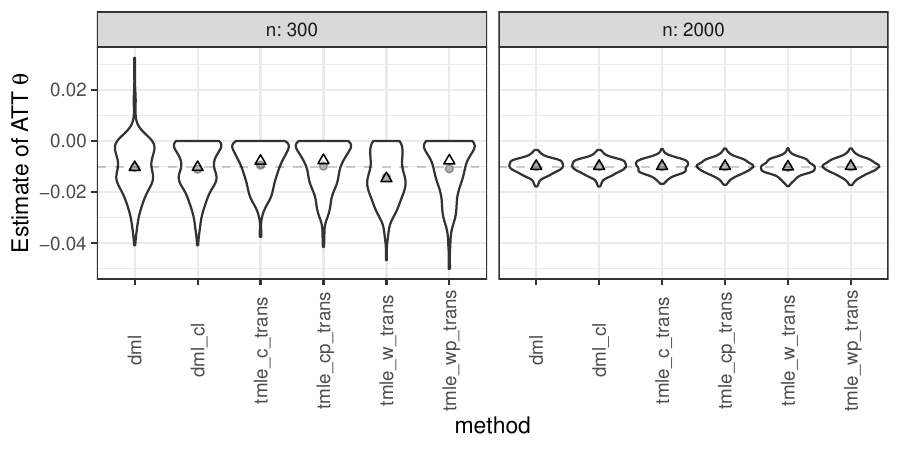}
    \caption{Six estimators' sampling distributions, where \texttt{trans} stands for the variant of TMLE in \citet{Balzer2016}. The bounds on $Q$ is set to be $[0,0.05]$. The horizontal dashed line is the truth. The gray dots and black triangles are the means and medians, respectively.}
    \label{fig: est small bound}
\end{figure}

\begin{figure}[bt]
    \centering
    \includegraphics[width=.9\linewidth]{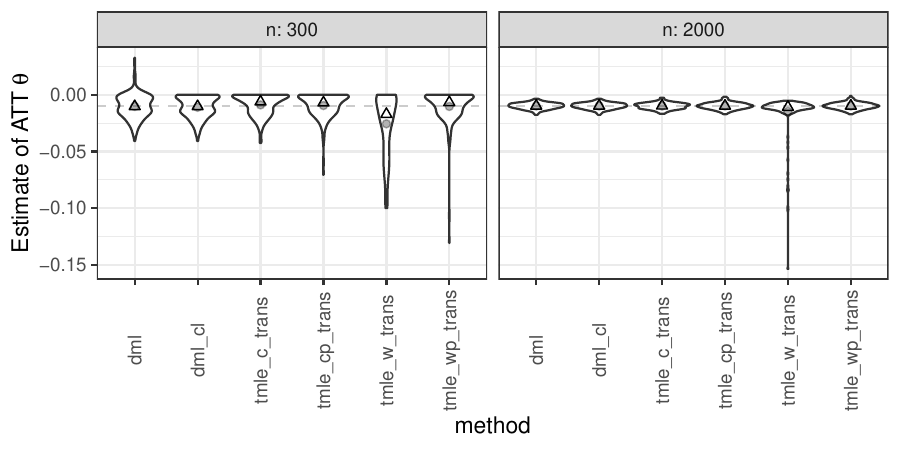}
    \caption{Figure similar to Fig.~\ref{fig: est small bound} with the bounds on $Q$ set to be $[0,0.2]$.}
    \label{fig: est large bound}
\end{figure}

\begin{figure}[bt]
    \centering
    \includegraphics[width=\linewidth]{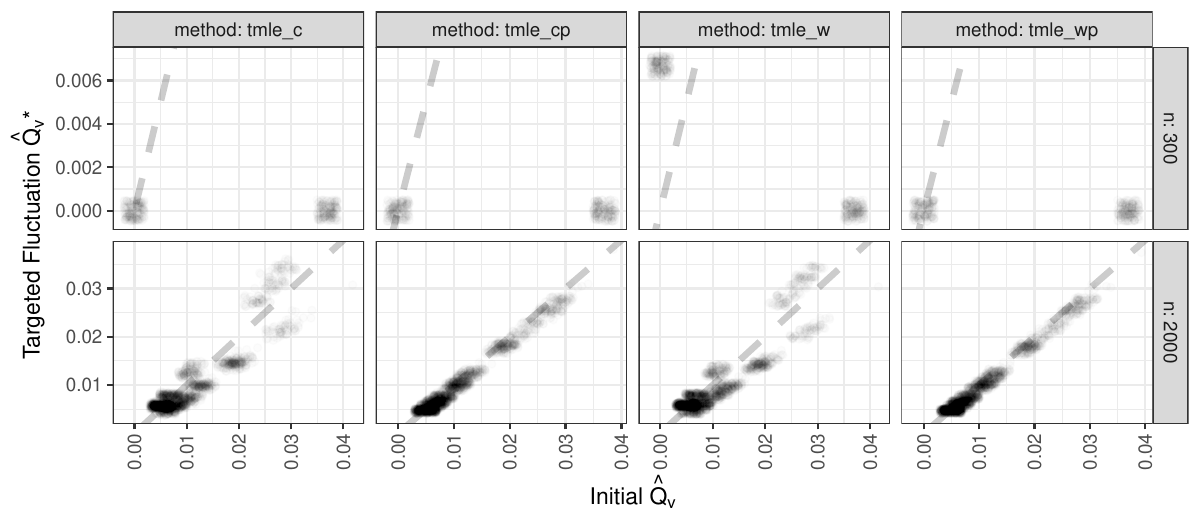}
    \caption{The fluctuated and the initial outcome regression estimators evaluated at each observed $X_i$ in two simulation runs. The points are jittered and semi-transparent for better visibility. The gray dashed line is the diagonal line $y=x$. Note the different y-axis scales between the two rows. The points appear far from the diagonal line in small samples ($n=300$, Row~1), indicating discordance between targeted fluctuation $\hat{Q}^*_v$ and initial $\hat{Q}_v$. In contrast, the points are closer to the diagonal line in relatively large samples ($n=2000$, Row~2), indicating more concordance.}
    \label{fig: Q}
\end{figure}

\end{document}